\documentclass[showpacs,preprint,showkeys]{revtex4}

\usepackage[dvips]{graphicx,color}
\usepackage{epsfig}
\usepackage{amsmath}
\usepackage{amssymb}
\usepackage{booktabs}

\allowdisplaybreaks[4]

\def\stau{\tilde{\tau}}
\def\neutralino{\tilde{\chi}}

\begin{document}

\title{Long life stau in the minimal supersymmetric standard model}

\author{Toshifumi Jittoh}
\email{jittoh@krishna.th.phy.saitama-u.ac.jp}
\author{Joe Sato}
\email{joe@phy.saitama-u.ac.jp}
\author{Takashi Shimomura}
\email{takashi@krishna.th.phy.saitama-u.ac.jp}
\author{Masato Yamanaka}
\email{masa@krishna.th.phy.saitama-u.ac.jp}
\affiliation{Department of Physics, Saitama University, 
        Shimo-okubo, Sakura-ku, Saitama, 338-8570, Japan}

\preprint{STUPP-05-183}



\begin{abstract}
 We study the stau lifetime in a scenario with the LSP taken to be a neutralino and the NLSP being a stau, based on the minimal
 supersymmetric Standard Model. The mass difference between the LSP and NLSP, $\delta m$, must 
 satisfy $\delta m/m_{\tilde{\chi}} \sim$ a few \% or less for coannihilation to occur, where
 $m_{\tilde{\chi}}$ is the neutralino mass. We calculate the stau lifetime from the decay modes 
 $\tilde{\tau}\rightarrow \tilde{\chi}\tau$, $\tilde{\chi}\nu_\tau\pi$, and 
 $\tilde{\chi}\nu_\tau\mu(e)\nu_{\mu(e)}$ and discuss its dependence on various parameters. We find that the
 lifetime is in the range $10^{-22}$--$10^{16}$ sec for $10^{-2} \le \delta m \le 10$ GeV. We also
 discuss the connection with lepton flavor violation if there is mixing between sleptons. 
\end{abstract}

\maketitle

\section{Introduction}
The existence of non-barionic dark matter is now confirmed and its density has been quantitatively 
estimated~\cite{Spergel:2003cb,Bennett:2003bz}. However its identity is still unknown. One of the most
prominent candidates is the weakly interacting massive particle
(WIMP)~\cite{Jungman:1995df,Bergstrom:2000pn,Munoz:2003gx,Bertone:2004pz}.

As is well known, the supersymmetric extension of the Standard Model provides a stable exotic particle,
the lightest supersymmetric particle (LSP), if R parity is conserved.  Among LSP candidates, the
neutralino LSP is the most suitable for non-barionic dark matter since its
nature fits that of the WIMP~\cite{Goldberg:1983nd,Ellis:1983ew}. Neutralinos are a linear
combination of the supersymmetric partners of the U(1) and SU(2) gauge bosons (bino and wino) and
the Higgs bosons (Higgsino).  They have mass in a range from 100 GeV to several TeV and are
electrically neutral. The lightest neutralino is stable if R parity is conserved.

Since the supersymmetric extension of the Standard Model is the most attractive theory, the nature of
neutralino dark matter has been studied
extensively~\cite{Ellis:1999mm}. 
In many scenarios of the supersymmetric model, the LSP neutralino consists mainly of the bino, the so-called
bino-like neutralino. In this case, naive calculations show that the predicted density in the
current universe is too high and it is necessary to find a way to reduce it. One mechanism to
suppress the density is coannihilation~\cite{Griest:1990kh}. If the next lightest supersymmetric particle
(NLSP) is nearly degenerate in mass with the LSP, the interaction of the LSP
with the NLSP is important in calculating the LSP annihilation process. For coannihilation to occur
tight degeneracy is necessary, since without coannihilation the LSP decouples from the thermal bath at 
$T \sim m/20$~\cite{Lee:1977ua}, where $m$ is the LSP mass. Therefore the mass difference $\delta m$
must satisfy $\delta m/m<$ a few \%, otherwise the NLSP decouples before coannihilation
becomes dominant. Furthermore, if the degeneracy is much tighter, we would observe a line spectrum of
photons from pair annihilation of dark matter~\cite{Hisano:2003ec,Matsumoto:2005ui}, since the
annihilation cross section of dark matter would be strongly enhanced due to the threshold correction. 

A candidate for the NLSP is the stau or stop in many class of MSSM, and in this paper we study the lifetime 
of the stau-like slepton having mass degenerate with the LSP neutralino. For the neutralino LSP to be dark matter, very
tight degeneracy is required in mass between the NLSP and the LSP neutralino. In particular the
heavier the LSP is, the tighter the degeneracy must be~\cite{Ellis:1999mm}. For such a degeneracy, the NLSP
is expected to have a long lifetime due to phase space suppression~\cite{Profumo:2004qt,Gladyshev:2005mn}.

An alternative scenario for a long-lived scalar particle is the gravitino LSP.  Considerable work has been devoted to
the long life NLSP in the context of the gravitino LSP. In this case, due to the
small coupling between a superWIMP (including the gravitino) and the NLSP, the lifetime of the NLSP becomes very 
long~\cite{Feng:2004mt,Hamaguchi:2004ne,Bi:2004ys}. To determine the most likely candidate for the LSP, we can accumulate and 
identify~\cite{Hamaguchi:2004df,Feng:2004yi} the candidate for long-lived NLSPs and 
compare the nature of the particles including couplings.

An implication on the fundamental feature of quantum mechanics would be brought by a long life stau. 
As noted in ref.~\cite{Jittoh:2004bz}, a Small-Q-value S-wave (SQS) decay can exhibit
non-exponential decay. A small Q value naively implies a small mass difference. For a gravitino 
LSP, the decay must occur in a P wave since the gravitino has spin 3/2. In contrast, in our case 
the decay can be S wave since all daughter particles have spin 0 or spin 1/2. Therefore, to extract
fundamental parameters from observations we must take special care in interpreting the 
results. This study offers the opportunity to examine a fundamental problem of quantum physics in
collider physics. Hence it is worthwhile studying the proposition that the stau is the NLSP and the lightest
neutralino is the LSP, and that their masses are tightly degenerate.

In Sec.~\ref{model}, we present the relevant Lagrangian and calculate the decay rate of the stau. In
Sec.~\ref{param}, we study the parameter dependences of the lifetime. We then consider the connection
with lepton flavor violation (LFV) in Sec.~\ref{lfv}. Finally we summarize our results in
Sec.~\ref{summary}. 

\section{Decay rate} \label{model}

In this section, we calculate the decay rate of the stau NLSP. The stau is a mass eigenstate consisting
of superpartneres of left- and right-handed taus,
\begin{equation}
 \stau = \cos \theta _{\tau} \stau_L + \sin \theta _{\tau} e^{-i\gamma_\tau}
 \stau_R.
\end{equation}
Here, $\theta _{\tau}$ is the mixing angle between $\tilde{\tau}_L$ and $\tilde{\tau}_R$, and 
$\gamma_{\tau}$ is the CP violating phase. The decay mode is governed by the mass difference, 
$\delta m \equiv m_{\text{NLSP}}-m_{\text{LSP}}$, according to kinematics. That is, the
lifetime of the stau depends strongly on $\delta m$.

\begin{figure}[ht]
\begin{center}
\includegraphics[height=30mm]{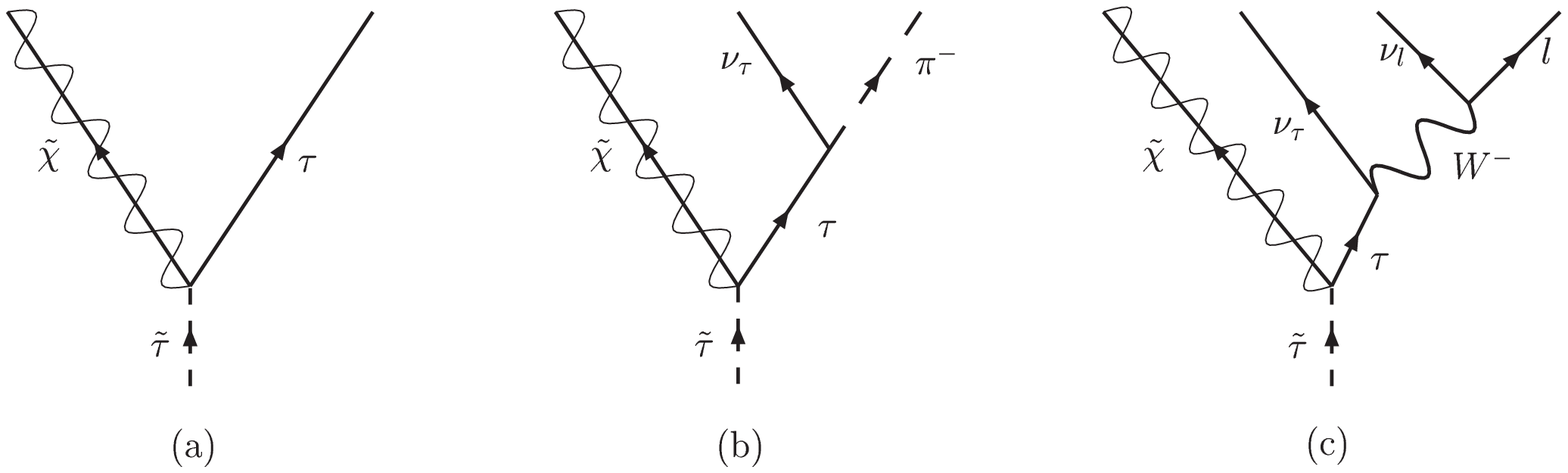}
\caption{Feynmann diagrams of stau decay: (a) $\stau \to \neutralino \tau$, \ (b) $\stau \to \neutralino \nu_\tau \pi$, \ (c) $\stau \to \neutralino l \nu_{\tau} \nu_l $.}
\label{diagrams}
\end{center}
\end{figure}

We consider the following four decay modes, $\stau \to \neutralino \tau$, 
$\stau \to \neutralino \nu_{\tau}\pi$, $\stau \to \neutralino \mu \nu_{\tau} \nu_{\mu} $, 
and $\stau \to \neutralino  e \nu_{\tau} \nu_e $ (see Fig.~\ref{diagrams}). Note that the NLSP can decay
into other particles, for example, if $\delta m > 1.86$ GeV, a D meson can be produced in the
stau decay but $\stau \to \neutralino \tau$ is dominant in this $\delta m$ region since the D meson
production process is suppressed by couplings and propagators.  In the 3-body and 4-body decay
processes, $\stau \to \neutralino \nu_{\tau} \pi$, $\stau \to \neutralino \mu \nu_{\tau} \nu_{\mu}$, 
and $\stau \to \neutralino e \nu_{\tau} \nu_e $, diagrams can be formulated with charginos as
intermediate states, however, such processes are strongly suppressed by the chargino propagator and we
can safely ignore them~\cite{Profumo:2004qt}.

In this paper, we consider the small mass difference case and hence we can ignore the momentum in the
W boson propagator. Thus the interaction Lagrangian is given by
\begin{equation}
\mathcal{L}_{int} = \stau ^* \bar{\neutralino} (g_LP_L+g_R P_R) \tau 
+ \frac{G}{\sqrt{2}} \nu_{\tau}\gamma_{\mu} P_L \tau J^{\mu} 
+ \frac{4 G}{\sqrt{2}}(\bar{l} \gamma ^{\mu} P_L \nu_{l})(\bar{\nu}_{\tau} \gamma _{\mu} P_L \tau)
+ h.c. \label{stau_lagrangian}
\end{equation}
The first term describes stau decay into a neutralino and a tau. Here, $P_L$ and $P_R$ are the
projection operators and $g_L$ and $g_R$ are the coupling constants given by, for example in
the bino-like neutralino case, 
\begin{equation}
g_L=\frac{g}{\sqrt{2} \cos{\theta _W}} \sin{\theta _W} \cos{\theta _{\tau}},\quad
g_R=\frac{\sqrt{2} g}{\cos{\theta _W}} \sin{\theta _W} \sin{\theta _{\tau}} e^{i \gamma _{\tau}},
\end{equation}
where $g$ is the weak coupling constant and $\theta _W$ is the Weinberg angle. The second and
third terms describe tau decays into a pion and/or leptons, where $G$ is the Fermi constant.

The detailed calculation of the stau decay rate is given in the appendix \ref{detail-calc}. Here we show
approximate formulae.  In the region $\delta m > m_\tau$, 2-body decay 
(see Fig.~\ref{diagrams}(a)) is allowed kinematically and is dominant.  The decay rate of the 
2-body final state is approximately
\begin{equation}
\Gamma_{\text{2-body}} = \frac{1}{4 \pi m_{\neutralino}} \sqrt{(\delta m)^2 - m_{\tau}^2}
\left( (g_L^2+|g_R|^2) \delta m -2 Re[g_Lg_R] m_{\tau} \right) ~, \label{approximated_Gamma2}
\end{equation}
where $m_{\stau}, m_{\neutralino}$, and $m_{\tau}$ are the masses of $\stau$, $\neutralino$, and $\tau$,
respectively. 

The 2-body decay is forbidden kinematically for $\delta m<m_\tau$, and pion production 
(see Fig.~\ref{diagrams}(b)) is dominant if $\delta m$ is larger than the pion mass $m_\pi$. 
The pion production 3-body decay rate has the approximate form
\begin{align}
\Gamma_{\text{3-body}}
&       = \frac{G^2 f_{\pi}^2 \cos ^2{\theta_c}}{210 (2 \pi )^3 m_{\neutralino} m_{\tau}^4} 
        \left( (\delta m)^2 - m_{\pi}^2 \right)^{5/2} \nonumber \\
&       \times\biggl[
        g_L^2 \delta m \left( 4(\delta m)^2 + 3 m_{\pi}^2 \right)
        - 2 Re[g_Lg_R]  m_{\tau} \left( 4(\delta m)^2 + 3 m_{\pi}^2 \right)
        +7 |g_R|^2 m_{\tau}^2 \delta m
        \biggr] ~.  \label{approximated_Gamma3}
\end{align}
Here $f_{\pi}$ is the pion decay constant and $\theta _c$ is the Cabbibo angle.

Incidentally, we note that a quark
cannot appear alone in any physical processes, a point that was missed in ref.~\cite{Profumo:2004qt}. Hence, $u$ and $d$ quarks appear only as mesons and
the $u, d$ production process is relevant only for $\delta m>m_\pi$.

Finally, when the mass difference is smaller than the pion mass, 4-body decay processes, 
$\stau \to \neutralino \mu \nu_{\tau} \nu_{\mu}$ and $ \stau \to \neutralino e \nu_{\tau} \nu_e$, 
are significant (see Fig.~\ref{diagrams}(c)). The approximate decay rate is calculated as
\begin{align}
\Gamma_{\text{4-body}}
&       = \frac{G^2}{945 (2 \pi )^5 m_{\neutralino} m_{\tau}^4} 
        \left( (\delta m)^2 - m_{l}^2 \right)^{5/2} \nonumber \\
&       \times\biggl[
        2 g_L^2 (\delta m)^3 \left( 2(\delta m)^2 - 19 m_{l}^2 \right)
        - 4 Re[g_Lg_R] m_{\tau} (\delta m)^2 \left( 2(\delta m)^2 - 19 m_{l}^2 \right) \nonumber \\
&       \qquad +3 |g_R|^2 m_{\tau}^2 \delta m \left( 2(\delta m)^2 - 23 m_{l}^2 \right)
        \biggr] ~.
\label{approximated_Gamma4}
\end{align}
Here $m_l$ is the charged lepton ($e$ or $\mu$) mass.

\section{Parameter dependence}\label{param}
In this section, we discuss the parameter dependence of the stau lifetime and the cosmological
constraints for the parameters in the bino-like LSP case. From Eqs.~(\ref{approximated_Gamma2}),
(\ref{approximated_Gamma3}), and (\ref{approximated_Gamma4}), we see that the stau lifetime depends
on $\delta m, \theta_{\tau}, m_{\neutralino}$, and $ \gamma_{\tau}$.  

\subsection{$m_{\tilde{\chi}}$}
First, we examine the neutralino mass dependence of the lifetime. We consider the cosmological
constraints for dark matter to get the mass range of the LSP. It is well known that the dark matter relic
density is reduced by the coannihilation process  and hence the neutralino mass can be heavier than it would be
without coannihilation. Accounting for the coannihilation process gives a neutralino mass
range~(the first ref. of \cite{Ellis:1999mm}) of 
\begin{equation}
 \begin{split}
  200~\text{GeV} \leq m_{\tilde {\chi }} \leq 600~\text{GeV}. \label{mass_range}
 \end{split}
\end{equation}
Here, we use CMSSM bound as a reference value of $m_{\tilde{\chi}}$, though we study the stau lifetime in general 
MSSM framework, since it is not strongly dependent on $m_{\tilde{\chi}}$ as noted below.
This is consistent with the cosmological constraint, $0.094 \leq \Omega_{DM}h^2 \leq 0.129$.

It is clear from Eqs.~\eqref{approximated_Gamma2}, \eqref{approximated_Gamma3}, and
\eqref{approximated_Gamma4} that the stau lifetime is proportional to the neutralino mass and
therefore in the mass range given in Eq.~\eqref{mass_range}, the stau lifetime varies by a
factor of three. Taking this into account, we use only $m_{\tilde \chi }=300 \text{GeV}$ in our figures.

\subsection{$\delta m$}
\begin{figure}[ht]
\begin{center}
\includegraphics[height=10cm]{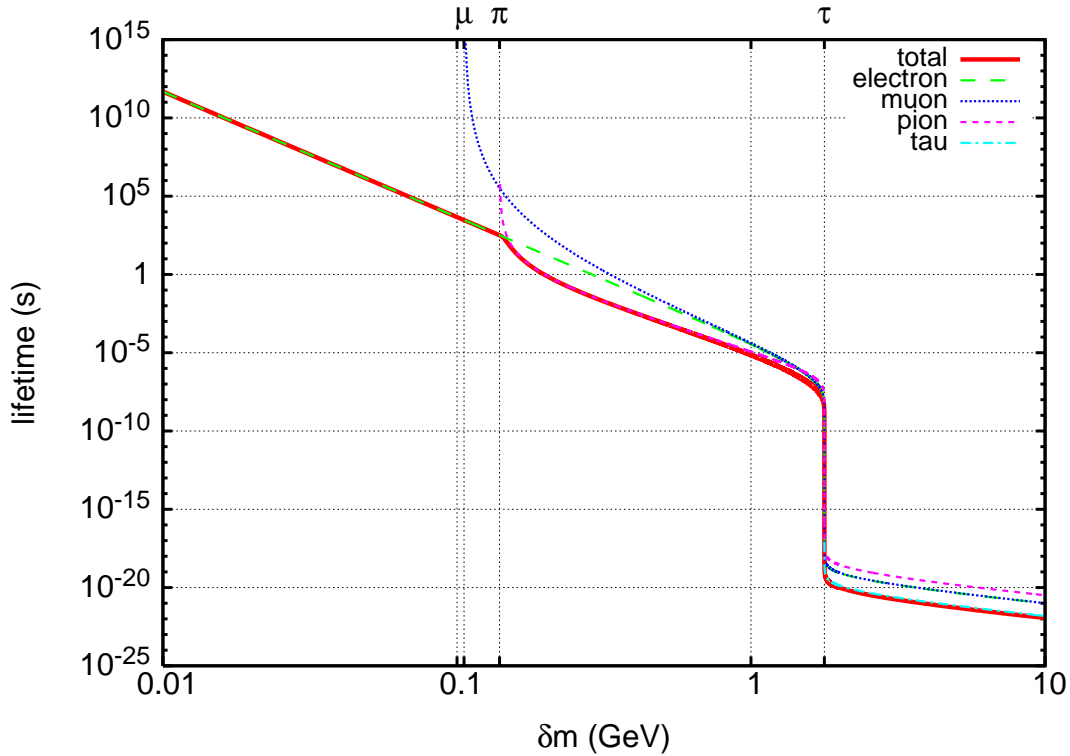}
\caption{Total lifetime and partial lifetimes of each decay
mode as a function of $\delta m$. The lines label electron, muon, pion, and
tau correspond to the processes $\stau \to \neutralino e \nu_{\tau} \nu_e$, $\stau \to \neutralino \mu \nu_{\tau} \nu_{\mu}$,
$\stau \to \neutralino \nu_{\tau} \pi$, and $\stau \to \neutralino
\tau$, respectively. Here we take $m_{\neutralino}=300$ GeV, $\theta _{\tau}=\pi /3$,
and $\gamma_{\tau}=0$.}
\label{lifetime300}
\end{center}
\end{figure}

Next, we consider the $\delta m$ dependence of the stau lifetime.  For coannihilation to occur,
the mass difference must satisfy $\delta m/m_{\neutralino} \sim$ a few \% or
smaller~\cite{Griest:1990kh}. Therefore, $\delta m$ must be smaller than a few GeV. The $\delta m$ dependence of the total stau lifetime and the partial lifetimes of
each decay mode in this 
$\delta m$ region is shown in Fig.~\ref{lifetime300}.  In Fig.~\ref{lifetime300}, we set the values as
follows: $m_{\neutralino}=300$ GeV, $\theta_{\tau}=\pi/3$, and $\gamma_{\tau}=0$.

>From Fig.~\ref{lifetime300}, we can see which mode is dominant for a certain $\delta m$.  The figure shows that the lifetime
increases drastically as $\delta m$ becomes smaller than the tau mass. This is because taus are produced in the region
$\delta m>m_{\tau}$, while in the region $\delta m < m_{\tau}$, taus cannot be
produced and instead pions appear in the final state. At $\delta m = m_{\pi}$ the lifetime
increases slightly. This is due to the fact that the dominant mode changes from 3-body to 4-body decay.
In contrast, at $\delta m = m_{\mu}$ the lifetime does not increase much, even though above this
mass, muons can 
be created. This is because at the pion mass, the muon production process is already kinematically
suppressed and the electron production process governs the stau decay.

To understand the $\delta m$ dependence of the lifetime quantitatively and intuitively we determine the
power of $\delta m$ in the decay rate, considering stau decay into neutralino
and $n-1$ massless particles.  The $\delta m$ dependence of the decay rate is determined by the phase
space and the squared amplitude~\cite{Profumo:2004qt}.

First, we examine the $\delta m$ dependence by considering the phase space. For 2-body decay, a phase space consideration gives
\begin{equation}
d \phi^{(2)} = \frac{d \Omega}{32\pi^2} 
\left( 1- \left( \frac{m_{\neutralino}}{m_{\neutralino}+\delta m} \right) ^2 \right)
\propto \delta m ~.
\end{equation}
By using a recursion relation between $d \phi ^{(n)}$ and $d \phi^{(n-1)}$, the phase space of $n$-body
decay renders the $\delta m$ dependence as
\begin{align}
d \phi ^{(n)} &\propto d \phi ^{(n-1)} \times \int^{\delta m} d \mu (d \phi^{(2)}) \nonumber \\
& \propto (\delta m)^{2(n-2)+1} ~.
\end{align}

Second, we consider the $\delta m$ dependence from the squared amplitude. If all of the $n-1$ massless
particles are fermions, the squared amplitude depends on $\delta m$ as
\begin{equation}
\mathcal{M}^{(n)} \propto (\delta m)^{n-1} ~,
\end{equation}
since it depends linearly on the massless fermion momentum. Thus, we obtain the dependence of the decay rate
on $\delta m$ for a final state of only fermions,
\begin{equation}
\Gamma^{(n)} \propto \mathcal{M}^{(n)} \times d \phi ^{(n)} \propto (\delta m)^{3n-4} \label{dm_dep_fermion} ~.
\end{equation}

In contrast, if one pion(NG-boson) appears in the stau decay process, the $\delta m$ dependence
of the squared amplitude becomes 
\begin{equation}
\mathcal{M}^{(n)} \propto (\delta m)^n ~.
\end{equation}
This change in the decay process is due to the fact that the amplitude of the pion production is proportional to the pion momentum. Namely,
the squared amplitude of the pion production process is proportional to the pion momentum squared.
Thus the $\delta m$ dependence of the process, in which one pion is involved, is
\begin{equation}
\Gamma^{(n)} \propto (\delta m)^{3(n-1)} \label{dm_dep_pi} ~.
\end{equation}
In the massless limit of external line particles, the $\delta m$ dependences of our results,
calculated in appendix \ref{detail-calc}, are 
\begin{align}
\Gamma_{\text{2-body}} &\propto (\delta m)^2, \nonumber \\
\Gamma_{\text{3-body}} &\propto (\delta m)^6, \nonumber \\
\Gamma_{\text{4-body}} &\propto (\delta m)^8 ~,
\end{align}
which are consistent with Eq.~(\ref{dm_dep_fermion}) and Eq.~(\ref{dm_dep_pi}).

More precisely, by taking into account the masses of the produced particles, we get the complicated 
$\delta m$ dependences 
\begin{align}
\Gamma_{\text{2-body}} &\propto (\delta m) \left( (\delta m)^2-m_{\tau}^2 \right)^{1/2}, \nonumber \\
\Gamma_{\text{3-body}} &\propto (\delta m) \left( (\delta m)^2-m_{\pi}^2 \right)^{5/2},\nonumber \\
\Gamma_{\text{4-body}} &\propto (\delta m)^3 \left( (\delta m)^2-m_{l}^2 \right)^{5/2}. \label{appro_thresh}
\end{align}
These equations clearly show the $\delta m$ dependence of the stau lifetime.

\subsection{$\gamma_\tau$}
\begin{figure}[h]
\begin{center}
\includegraphics{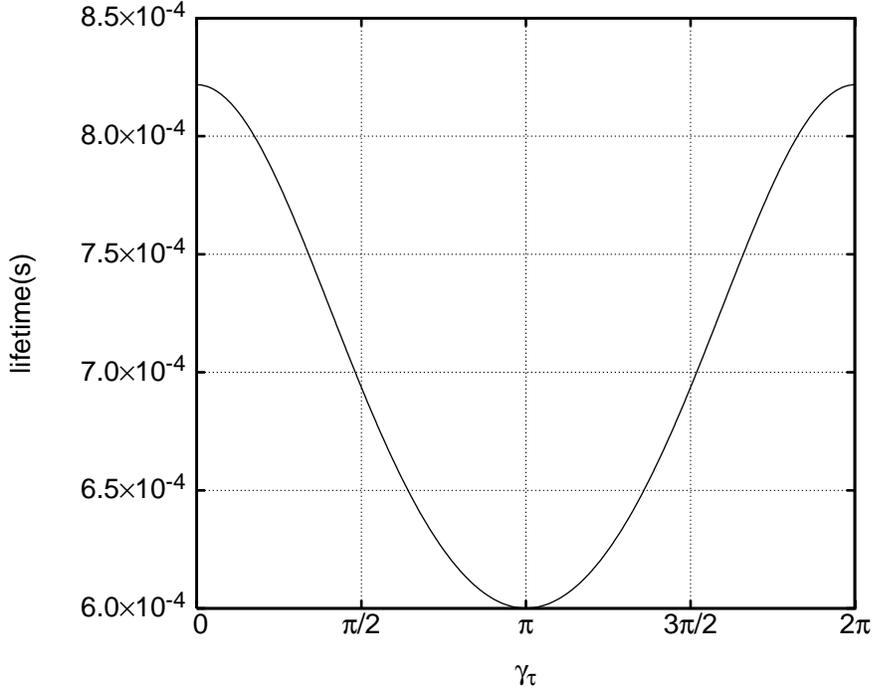}
\caption{CP violating phase and stau lifetime for 
$m_{\tilde \chi }=300 \text{GeV}, \delta m=0.5 \text{GeV}$, and $\theta _\tau =\pi /3$.}
\label{cp_life}
\end{center}
\end{figure}

We next consider the $\gamma_\tau$ dependence of the stau lifetime.
Figure~\ref{cp_life} shows the lifetime as a function of the CP violating phase.
We set the other parameters to be $m_{\tilde \chi }=300 \text{GeV}$, 
$\delta m=0.5 \text{GeV}$, 
and $\theta _\tau =\pi /3$.

>From Fig.~\ref{cp_life}, it is clear that the CP violating phase does not greatly affect the stau
lifetime, and so we fix $\gamma _\tau =0$ (no CP violation).

As expressed in 
Eq.~\eqref{approximated_Gamma4}, the effect of CP violation appears in the $Re[g_Lg_R]$ terms only. Since
the coefficients of the $Re[g_Lg_R]$ terms are smaller than those of $|g_R|^2$, it is again clear that
the CP violating phase does not greatly affect the stau lifetime. 

\subsection{$\theta_\tau$}
\begin{figure}[ht]
\includegraphics[height=55mm]{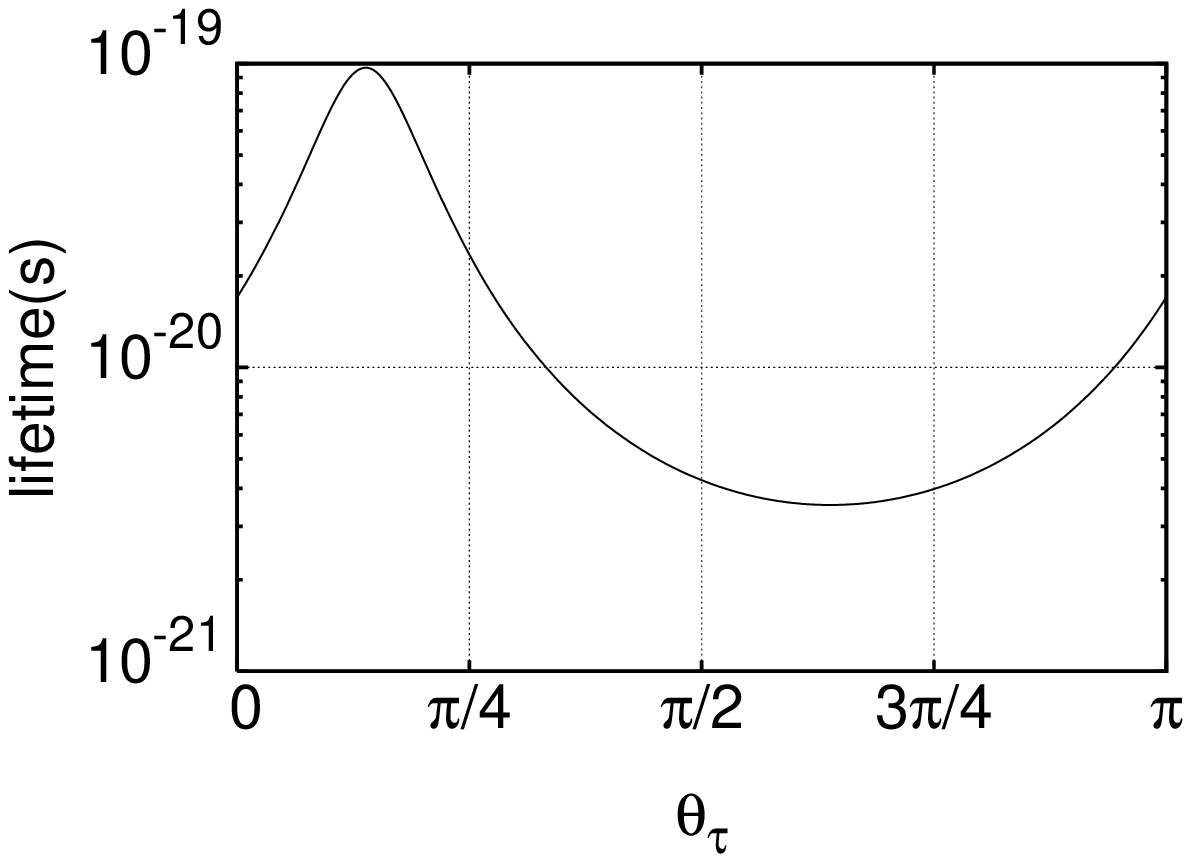}
\includegraphics[height=55mm]{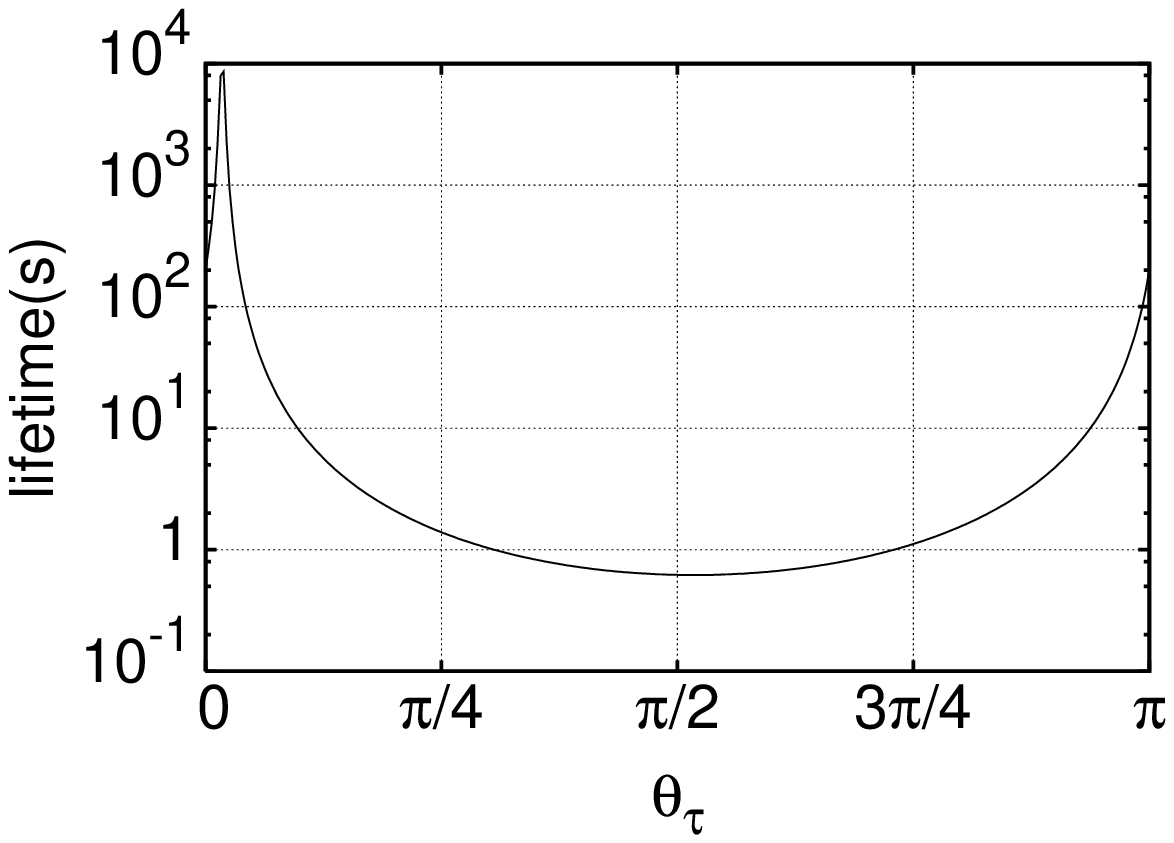}
\caption{Stau mixing angle and stau lifetime with $m_{\tilde \chi }=300 \text{GeV}$ and 
$\gamma _\tau =0$. Left panel: $\delta m = 2.0$ GeV, right panel: $\delta m=0.2$.}
\label{enlargedlifetime}
\end{figure}

The $\theta _\tau $ dependence of the stau lifetime is as strong as the $\delta m$ dependence. Figure~\ref{enlargedlifetime} shows the $\theta_\tau$ dependence of the stau 
lifetime with $m_{\tilde \chi }=300 \text{GeV}$ and $\gamma _\tau =0$.

We can see from the right panel in Fig.~\ref{enlargedlifetime} that for
$\delta m \ll m_\tau $, $\tilde \tau _R$ decays much more quickly than $\tilde \tau _L$.
This can be understood by considering two steps. First, we note that only left-handed virtual taus contribute to the
final state $\nu_\tau$. Second, $\tilde \tau _R$ converts to $\tau_L$ by picking up $m_\tau$ in the
tau propagator, while $\tilde \tau _L$ converts by picking up the momentum $p_\tau$ in the propagator.
Since $p_\tau \sim \delta m \ll m_\tau$, the former contribution is much larger and hence there is a
strong dependence on $\theta_\tau$. 

\subsection{Constraint from Big Bang Nucleosynthesis}
Finally, we consider the constraints from Big Bang nucleosynthesis (BBN). The stau lifetime will be constrained by
the standard BBN scenario. Particularly, in the case where a stau decays into hadrons, the constraint
on the stau lifetime is more stringent~\cite{Kawasaki:2000qr,Kawasaki:2004qu}. At $t \geq 100$ (sec)
($t$ is the age of the universe), pions emitted from a stau decay also decay before they interact with the
background nucleons, so they do not affect BBN. However, if pions are emitted at $t \leq  100$
(sec), they inter-convert a background proton and neutron into each other, even after the normal
freeze-out time of the n/p ratio. Therefore we must take careful consideration to the parameter region in
which pions are emitted for (BBN start) $\leq t \leq$ (BBN end), i.e. 
$10^{-2} \text{(sec)}\leq t \leq 10^2 \text{(sec)}$. Naively interpreting Fig.~\ 38 in
ref.~\ \cite{Kawasaki:2004qu}, we can conclude that the corresponding parameter region can neglect the
constraints from BBN. This is inferred from the fact that the $\tilde \tau$ number density is roughly two
orders lower than that of baryons, $Y_{\tilde \tau} \sim 10^{-12}$, since $m_{\tilde \chi}$ is of the order of
several hundreds of GeV, and the released energy is less than $1$ GeV. Here $Y_{\tilde \tau}$ is the ratio
of the stau number density to the entropy density. Thus $E_{\text vis}Y_{\tilde \tau} < 10^{-12}$ GeV,
which does not influence the BBN for $t < 10^{2}$ sec.
Incidentally, it is apparent from Fig.5 that stau lifetime can not exceed the age of universe (about 4.1 $\times 10^
{17}$ sec). It is consistent with the fact that an exotic heavy charged particle has not been observed yet.
\begin{figure}[ht]
\includegraphics[height=80mm]{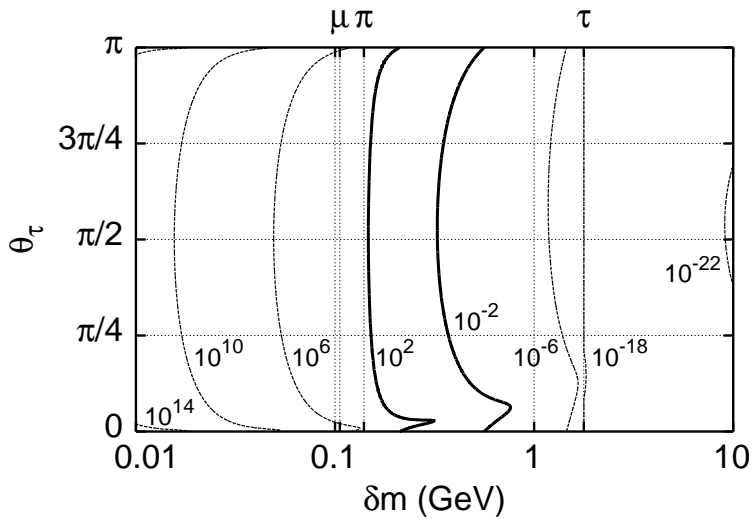}
\caption{Contour plot of stau lifetime in sec as a function of $\delta m$ and $\theta_\tau$.
To emphasize the BBN era, we draw thick lines for $10^{-2}$ and $10^2$ sec.}
\label{allowed_region}
\end{figure}

\section{Connection with Lepton Flavor Violation}\label{lfv}
We next consider LFV\footnote{For the ref. of superWIMP case, see for
example\cite{Hamaguchi:2004ne}}. The NLSP slepton might be a linear combination of
flavor eigenstates. It is expected that lepton flavor violating events will be observed due to this mixing,
such as $\mu\rightarrow e\gamma$. If we observe $\tau\rightarrow e\gamma$ or $\tau\rightarrow
\mu\gamma$ events, then within the context of the minimal supersymmetric Standard Model (MSSM), we
would conclude that the selectron forms part of the NLSP:
\begin{eqnarray}
 \phi_{\rm NLSP}=N_1 \tilde{e} + \sqrt{1-N_1^2} \tilde{\tau} .
\label{eq:LFVNLSP}
\end{eqnarray}
The branching ratio of $\tau\rightarrow e\gamma$ is roughly proportional to $N_1^2$. The current
upper bound on the branching ratio, $<O(10^{-7})$, gives a poor constraint, at most $N_1<0.1$. If
$N_1\ne 0$, the NLSP slepton can decay, 
\begin{eqnarray}
 \phi_{\rm NLSP}\rightarrow\tilde{\chi}+e,
\end{eqnarray}
with a lifetime of
\begin{eqnarray}
1.666 \times 10^{-18} \left( \frac{m_{\neutralino}}{300{\rm GeV}} \right) 
\left( \frac{1.0{\rm GeV}}{\delta m} \right)^2 \left( \frac{0.1}{N_1} \right)^2 {\rm sec}~.
\end{eqnarray}
We may interpret the result as implying that the NLSP slepton consists only of a scalar electron since it decays
only into an electron. The lifetime is so short, even if $\delta m < m_\tau$, that we cannot
accumulate the NLSP. However if we can measure the lifetime, we can determine the fraction of the scalar 
electron in the NLSP slepton.  Further evidence is given by flavor violating processes of
both charged leptons~\cite{Hisano:1998fj,Hisano:1995cp,Hisano:1995nq,Sato:2000ff} and
the neutrino~\cite{Ota:2005et}. It is important to interpret all the LFV processes together. 

\section{summary}\label{summary}
We have studied an MSSM scenario in which the LSP and NLSP are a bino-like neutralino and a stau,
respectively. Since the mass difference is, in many cases, assumed to be degenerate, from the
requirement of coannihilation, we paid special attention to the very small $\delta m$ case.
We calculated the partial lifetimes for the decay modes shown in Fig.~\ref{diagrams}.

We have investigated the stau lifetime dependence on $\delta m$, $\theta_\tau$, $\gamma_\tau$,
and $m_{\tilde{\chi}}$, considering cosmological constraints. The lifetime strongly depends on $\delta m$ and
$\theta_\tau$, while it is almost independent of $\gamma_\tau$ and $m_{\tilde{\chi}}$.

The stau lifetime dependence on $\delta m$ changes as each threshold is crossed. When $\delta m$ is
larger than $m_\tau$, the lifetime increases in proportion to $(\delta m)^{-2}$ as $\delta m$
decreases. In the range $m_\tau>\delta m>m_\pi$ the lifetime obeys the scaling 
$\sim (\delta m)^{-6}$. Below $m_\pi$, it grows with $(\delta m)^{-8}$. The $\delta m$
dependence of the stau lifetime can be largely understood by counting the mass dimension of phase space
and the squared amplitude in the massless limit of Standard Model particles. While the massless limit is
a good approximation in regions far from the thresholds, the $\delta m$ dependence near the
thresholds are given by Eq.~\eqref{appro_thresh}. 

The lifetime also strongly depends on $\theta_\tau$, as shown in
Fig.~\ref{enlargedlifetime}. $\tilde{\tau}_R$ contributes to 3- and 4-body decay processes by
picking up the $m_\tau$ term in the intermediate $\tau$ propagator, while $\tilde{\tau}_L$ picks up the
$p_\tau$ term. Since $p_\tau \sim \delta m \ll m_\tau$, the contribution for $\tilde{\tau}_R$
is much larger and hence there is a strong dependence on $\theta_\tau$.
 
As is seen in Fig.~\ref{lifetime300}, if $\delta m$ is smaller than $m_\tau$,  a stau can be very 
long-lived. This fact may need to be taken into account in studies at the Large Hadron Collider
(LHC) and the International Linear Collider (ILC). The
superWIMP case is similar. However, in the gravitino scenario an energetic $\tau$ is apparently produced, while in
our scenario we should observe very low energy $\pi$, $\mu$, or $e$. Therefore a clean
experiment is required. The ILC would be most suitable for investigating the nature of the NLSP slepton.

Another candidate for the NLSP is the stop. However, it is more complicated to investigate this
possibility. A stop must always be accompanied by quark(s); it forms a meson-like fermion with one quark or
a baryon-like boson with two quarks. It is almost impossible to calculate the exact mass eigenvalue
using QCD. It is beyond the scope of this paper and is left for future work.

We have also discussed lepton flavor violation due to slepton mixing. If there is even a tiny
component of a scalar electron or a scalar muon in the NLSP ``stau'', the decay signal of the NLSP will be 
completely different from the pure stau case. The NLSP slepton undergoes 2-body decay into the accompanying
electron or muon. Since it is a 2-body process, it occurs very quickly, $\sim ~(10^{-20})N_1^{-2}$
sec where $N_1$ represents the portion of the scalar electron or scalar muon, as shown in
Eq.~(\ref{eq:LFVNLSP}).  As this mixing causes charged/neutral lepton flavor violation, it is very
important to compare the NLSP slepton decay with other processes such as $\tau\rightarrow e(\mu)\gamma$.
To fully clarify the nature of the NLSP, we need to interpret all the LFV processes together.

SQS decay takes place in a case of small mass difference. It will appear strongly, if decay rate is small. 
Since long lifetime means small decay rate, long-lived stau will give a good chance to observe a non-exponential decay.
In contrast, if there is no LFV in slepton mixing, the pure stau has a very long lifetime
and it is then possible to experimentally observe SQS decay~\cite{Jittoh:2004bz}. This is
important to check the fundamental feature of quantum mechanics. 

Hence, the small $\delta m$ case is very interesting and its study is very important.

\begin{acknowledgments}
 J.~S. would like to thank K.~Hamaguchi for his useful comments. The work of J.~S. is partially
 supported by a Grant-in-Aid for Scientific Research on Priority Areas No.~16028202 and No.~17740131. 
\end{acknowledgments}

\appendix
\section{}\label{detail-calc}
In this appendix, we show the exact and approximate stau decay rates. We use the exact decay rate in
formulating the figures. We make the following approximations: we keep only the leading order term of 
$(\delta m/m_{\tilde{\chi}})$ and we replace the denominator of the $\tau$ propagator by $m_\tau^2$ in
the 3- and 4-body cases. We calculate the decay rates of the three processes shown in
Fig.~\ref{diagrams} with the Lagrangian of Eq.~(\ref{stau_lagrangian}). 

\subsection*{2-body decay}
The decay rate of the 2-body decay process (see Fig.~\ref{diagrams}(a)) is given by
\begin{align}
\Gamma_{\text{2-body}} = \frac{1}{16 \pi m_{\stau}^3} &(m_{\stau}^4 +
m_{\neutralino}^4 + m_{\tau}^4 - 2 m_{\stau}^2 m_{\neutralino}^2 - 2
m_{\stau}^2 m_{\tau}^2 - 2 m_{\neutralino}^2 m_{\tau}^2) ^{\frac{1}{2}}
\nonumber \\ \times &\Bigl\{
(g_L^2+|g_R|^2)(m_{\stau}^2-m_{\neutralino}^2-m_{\tau}^2) - 4 Re[g_Lg_R]
m_{\tau} m_{\neutralino} \Bigr\} \label{Gamma2} ~.
\end{align} 

For the analysis discussed in Sec.~\ref{model}, we approximate the decay rate as 
\begin{equation}
\Gamma_{\text{2-body}} = \frac{1}{4 \pi m_{\neutralino}} \sqrt{(\delta m)^2 -
m_{\tau}^2} \left( (g_L^2+|g_R|^2) \delta m -2Re[g_Lg_R]m_{\tau}
\right) ~.
\end{equation}

\subsection*{3-body decay}
The decay rate of the 3-body decay process (see Fig.~\ref{diagrams}(b)) is calculated as
\begin{align}
 \Gamma_{\text{3-body}} 
 = 
 &\frac{G^2 f_{\pi}^2 \cos ^2{\theta_c} \left( (\delta m)^2-m_{\pi}^2 \right)}{64 \pi ^3 m_{\stau}^{3}} \nonumber \\
 &\times
	\int _0^1
	dx
	\sqrt{\left( (\delta m)^2-q_f^2 \right) \left( (\delta m+2m_{\neutralino})^2-q_f^2 \right)}
	~ \frac{1}{(q_f^2-m_{\tau}^2)^2 + (m_{\tau} \Gamma_{\tau})^2} \nonumber \\
&	\times
	(q_f^2 - m_{\pi}^2)
	\Biggl[
 	\frac{1}{4}(g_L^2 q_f^2 + |g_R^2| m_{\tau}^2) ((\delta m)^2 + 2m_{\neutralino} \delta m - q_f^2)
	- Re[g_Lg_R] m_{\neutralino} m_{\tau} q_f^2 
	\Biggr]  \label{Gamma3} ~.
\end{align}
Here $q_f^2$ is given as
\begin{equation}
q_f^2=(\delta m)^2- \left( (\delta m)^2-m_f^2 \right) x \label{qf} ~,
\end{equation}
where the index $f (=\pi, e, \mu)$ denotes a massive particle, except the neutralino, in the final states; 
$f=\pi$ in the 3-body case. $\Gamma_\tau$ is the tau decay width and $(m_\tau \Gamma_\tau)^2$ 
is added to the denominator of the tau propagator for the region $\delta m \ge m_\tau$.

The approximate decay rate is 
\begin{align}
\Gamma_{\text{3-body}}
&	= \frac{G^2 f_{\pi}^2 \cos ^2{\theta_c}}{210 (2 \pi )^3 m_{\neutralino} m_{\tau}^4} 
	\left( (\delta m)^2 - m_{\pi}^2 \right)^{5/2} \nonumber \\
&	\times\biggl[
	g_L^2 \delta m \left( 4(\delta m)^2 + 3 m_{\pi}^2 \right)
	- 2 Re[g_Lg_R]  m_{\tau} \left( 4(\delta m)^2 + 3 m_{\pi}^2 \right)
	+7 |g_R|^2 m_{\tau}^2 \delta m
	\biggr] ~.
\end{align}

\subsection*{4-body decay}
In the 4-body decay processes (see Fig.~\ref{diagrams}(c)), the decay rate is given by
\begin{align}
	\Gamma_{\text{4-body}} 
	= 
&	\frac{G^2 \left( (\delta m)^2-m_l^2 \right)}{24 (2 \pi) ^5 m_{\stau}^{3}} \nonumber \\
&	\times \int _0^1
	dx
	\sqrt{\left( (\delta m)^2-q_f^2 \right) \left( (\delta m+2m_{\neutralino})^2-q_f^2 \right) }
	~ \frac{1}{(q_f^2-m_{\tau}^2)^2 + (m_{\tau} \Gamma_{\tau})^2}
	\frac{1}{q_f^4} \nonumber \\
&	\times \Biggl[
	\Biggl\{
	\frac{1}{4}(g_L^2 q_f^2 + |g_R^2| m_{\tau}^2) ((\delta m)^2 + 2m_{\neutralino} \delta m - q_f^2)
	- Re[g_Lg_R] m_{\neutralino} m_{\tau} q_f^2
	\Biggr\} \nonumber \\
&	\times
	\Biggl\{
	12 m_l^4 q_f^4 \log \left[ \frac{q_f^2}{m_l^2} \right]
	+ (q_f^4-m_l^4) (q_f^4 -8 m_l^2 q_f^2 + m_l^4)
	\Biggr\}
	\Biggr], \label{Gamma4} 
\end{align}
where $l=e, \mu$ and $q_l^2$ is given by Eq.~(\ref{qf}).

We can approximate the decay rate as
\begin{align}
\Gamma_{\text{4-body}}
&	= \frac{G^2}{945 (2 \pi )^5 m_{\neutralino} m_{\tau}^4} 
	\left( (\delta m)^2 - m_{l}^2 \right)^{5/2} \nonumber \\
&	\times\biggl[
	2 g_L^2 (\delta m)^3 \left( 2(\delta m)^2 - 19 m_{l}^2 \right)
	- 4 Re[g_Lg_R] m_{\tau} (\delta m)^2 \left( 2(\delta m)^2 - 19 m_{l}^2 \right) \nonumber \\
&	\qquad +3 |g_R|^2 m_{\tau}^2 \delta m \left( 2(\delta m)^2 - 23 m_{l}^2 \right)
	\biggr] ~.
\end{align}

\bibliographystyle{apsrev}

\begin{thebibliography}{35}
\expandafter\ifx\csname natexlab\endcsname\relax\def\natexlab#1{#1}\fi
\expandafter\ifx\csname bibnamefont\endcsname\relax
  \def\bibnamefont#1{#1}\fi
\expandafter\ifx\csname bibfnamefont\endcsname\relax
  \def\bibfnamefont#1{#1}\fi
\expandafter\ifx\csname citenamefont\endcsname\relax
  \def\citenamefont#1{#1}\fi
\expandafter\ifx\csname url\endcsname\relax
  \def\url#1{\texttt{#1}}\fi
\expandafter\ifx\csname urlprefix\endcsname\relax\def\urlprefix{URL }\fi
\providecommand{\bibinfo}[2]{#2}
\providecommand{\eprint}[2][]{\url{#2}}


\bibitem[{\citenamefont{Spergel et~al.}(2003)}]{Spergel:2003cb}
\bibinfo{author}{\bibfnamefont{D.~N.} \bibnamefont{Spergel}}
  \bibnamefont{et~al.} (\bibinfo{collaboration}{WMAP}),
  \bibinfo{journal}{Astrophys. J. Suppl.} \textbf{\bibinfo{volume}{148}},
  \bibinfo{pages}{175} (\bibinfo{year}{2003}), \eprint{astro-ph/0302209}.

\bibitem[{\citenamefont{Bennett et~al.}(2003)}]{Bennett:2003bz}
\bibinfo{author}{\bibfnamefont{C.~L.} \bibnamefont{Bennett}}
  \bibnamefont{et~al.}, \bibinfo{journal}{Astrophys. J. Suppl.}
  \textbf{\bibinfo{volume}{148}}, \bibinfo{pages}{1} (\bibinfo{year}{2003}),
  \eprint{astro-ph/0302207}.

\bibitem[{\citenamefont{Jungman et~al.}(1996)\citenamefont{Jungman,
  Kamionkowski, and Griest}}]{Jungman:1995df}
\bibinfo{author}{\bibfnamefont{G.}~\bibnamefont{Jungman}},
  \bibinfo{author}{\bibfnamefont{M.}~\bibnamefont{Kamionkowski}},
  \bibnamefont{and} \bibinfo{author}{\bibfnamefont{K.}~\bibnamefont{Griest}},
  \bibinfo{journal}{Phys. Rept.} \textbf{\bibinfo{volume}{267}},
  \bibinfo{pages}{195} (\bibinfo{year}{1996}), \eprint{hep-ph/9506380}.

\bibitem[{\citenamefont{Bergstrom}(2000)}]{Bergstrom:2000pn}
\bibinfo{author}{\bibfnamefont{L.}~\bibnamefont{Bergstrom}},
  \bibinfo{journal}{Rept. Prog. Phys.} \textbf{\bibinfo{volume}{63}},
  \bibinfo{pages}{793} (\bibinfo{year}{2000}), \eprint{hep-ph/0002126}.

\bibitem[{\citenamefont{Munoz}(2004)}]{Munoz:2003gx}
\bibinfo{author}{\bibfnamefont{C.}~\bibnamefont{Munoz}}, \bibinfo{journal}{Int.
  J. Mod. Phys.} \textbf{\bibinfo{volume}{A19}}, \bibinfo{pages}{3093}
  (\bibinfo{year}{2004}), \eprint{hep-ph/0309346}.

\bibitem[{\citenamefont{Bertone et~al.}(2005)\citenamefont{Bertone, Hooper, and
  Silk}}]{Bertone:2004pz}
\bibinfo{author}{\bibfnamefont{G.}~\bibnamefont{Bertone}},
  \bibinfo{author}{\bibfnamefont{D.}~\bibnamefont{Hooper}}, \bibnamefont{and}
  \bibinfo{author}{\bibfnamefont{J.}~\bibnamefont{Silk}},
  \bibinfo{journal}{Phys. Rept.} \textbf{\bibinfo{volume}{405}},
  \bibinfo{pages}{279} (\bibinfo{year}{2005}), \bibinfo{note}{for review},
  \eprint{hep-ph/0404175}.

\bibitem[{\citenamefont{Goldberg}(1983)}]{Goldberg:1983nd}
\bibinfo{author}{\bibfnamefont{H.}~\bibnamefont{Goldberg}},
  \bibinfo{journal}{Phys. Rev. Lett.} \textbf{\bibinfo{volume}{50}},
  \bibinfo{pages}{1419} (\bibinfo{year}{1983}).

\bibitem[{\citenamefont{Ellis et~al.}(1984)\citenamefont{Ellis, Hagelin,
  Nanopoulos, Olive, and Srednicki}}]{Ellis:1983ew}
\bibinfo{author}{\bibfnamefont{J.~R.} \bibnamefont{Ellis}},
  \bibinfo{author}{\bibfnamefont{J.~S.} \bibnamefont{Hagelin}},
  \bibinfo{author}{\bibfnamefont{D.~V.} \bibnamefont{Nanopoulos}},
  \bibinfo{author}{\bibfnamefont{K.~A.} \bibnamefont{Olive}}, \bibnamefont{and}
  \bibinfo{author}{\bibfnamefont{M.}~\bibnamefont{Srednicki}},
  \bibinfo{journal}{Nucl. Phys.} \textbf{\bibinfo{volume}{B238}},
  \bibinfo{pages}{453} (\bibinfo{year}{1984}).




\bibitem[{\citenamefont{Ellis et~al.}(2000)\citenamefont{Ellis, Falk, Olive,
  and Srednicki}}]{Ellis:1999mm}
\bibinfo{author}{\bibfnamefont{J.~R.} \bibnamefont{Ellis}},
  \bibinfo{author}{\bibfnamefont{T.}~\bibnamefont{Falk}},
  \bibinfo{author}{\bibfnamefont{K.~A.} \bibnamefont{Olive}}, \bibnamefont{and}
  \bibinfo{author}{\bibfnamefont{M.}~\bibnamefont{Srednicki}},
  \bibinfo{journal}{Astropart. Phys.} \textbf{\bibinfo{volume}{13}},
  \bibinfo{pages}{181} (\bibinfo{year}{2000}), \eprint{hep-ph/9905481}.

\bibinfo{author}{\bibfnamefont{J.~R.} \bibnamefont{Ellis}},
  \bibinfo{author}{\bibfnamefont{K.~A.} \bibnamefont{Olive}}, \bibnamefont{and}
  \bibinfo{author}{\bibfnamefont{Y.}~\bibnamefont{Santoso}},
  \bibinfo{journal}{Astropart. Phys.} \textbf{\bibinfo{volume}{18}},
  \bibinfo{pages}{395} (\bibinfo{year}{2003}), \eprint{hep-ph/0112113}.

\bibinfo{author}{\bibfnamefont{J.}~\bibnamefont{Edsjo}} \bibnamefont{and}
  \bibinfo{author}{\bibfnamefont{P.}~\bibnamefont{Gondolo}},
  \bibinfo{journal}{Phys. Rev.} \textbf{\bibinfo{volume}{D56}},
  \bibinfo{pages}{1879} (\bibinfo{year}{1997}), \eprint{hep-ph/9704361}.

\bibinfo{author}{\bibfnamefont{T.}~\bibnamefont{Nihei}} \bibnamefont{and}
  \bibinfo{author}{\bibfnamefont{M.}~\bibnamefont{Sasagawa}},
  \bibinfo{journal}{Phys. Rev.} \textbf{\bibinfo{volume}{D70}},
  \bibinfo{pages}{055011} (\bibinfo{year}{2004}), \eprint{hep-ph/0404100}.

\bibinfo{author}{\bibfnamefont{K.~A.} \bibnamefont{Olive}}
  (\bibinfo{year}{2002}), \eprint{astro-ph/0202486}.

  M.~M.~Nojiri,
  Pramana {\bf 62}, 335 (2004)
  [arXiv:hep-ph/0305192].

  K.~Kohri, M.~Yamaguchi and J.~Yokoyama,
  Phys.\ Rev.\ D {\bf 72}, 083510 (2005)
  [arXiv:hep-ph/0502211].

\bibinfo{author}{\bibfnamefont{M.}~\bibnamefont{Drees}}, \bibinfo{journal}{AIP
  Conf. Proc.} \textbf{\bibinfo{volume}{805}}, \bibinfo{pages}{48}
  (\bibinfo{year}{2006}), \eprint{hep-ph/0509105}.

\bibinfo{author}{\bibfnamefont{H.}~\bibnamefont{Baer}},
  \bibinfo{author}{\bibfnamefont{A.}~\bibnamefont{Mustafayev}},
  \bibinfo{author}{\bibfnamefont{E.-K.} \bibnamefont{Park}}, \bibnamefont{and}
  \bibinfo{author}{\bibfnamefont{S.}~\bibnamefont{Profumo}},
  \bibinfo{journal}{JHEP} \textbf{\bibinfo{volume}{07}}, \bibinfo{pages}{046}
  (\bibinfo{year}{2005}), \eprint{hep-ph/0505227}.

\bibinfo{author}{\bibfnamefont{U.}~\bibnamefont{Chattopadhyay}},
  \bibinfo{author}{\bibfnamefont{D.}~\bibnamefont{Choudhury}},
  \bibinfo{author}{\bibfnamefont{M.}~\bibnamefont{Drees}},
  \bibinfo{author}{\bibfnamefont{P.}~\bibnamefont{Konar}}, \bibnamefont{and}
  \bibinfo{author}{\bibfnamefont{D.~P.} \bibnamefont{Roy}},
  \bibinfo{journal}{Phys. Lett.} \textbf{\bibinfo{volume}{B632}},
  \bibinfo{pages}{114} (\bibinfo{year}{2006}), \eprint{hep-ph/0508098}.

\bibinfo{author}{\bibfnamefont{M.}~\bibnamefont{Drees}},
  \bibinfo{author}{\bibfnamefont{M.~M.} \bibnamefont{Nojiri}},
  \bibinfo{author}{\bibfnamefont{D.~P.} \bibnamefont{Roy}}, \bibnamefont{and}
  \bibinfo{author}{\bibfnamefont{Y.}~\bibnamefont{Yamada}},
  \bibinfo{journal}{Phys. Rev.} \textbf{\bibinfo{volume}{D56}},
  \bibinfo{pages}{276} (\bibinfo{year}{1997}), \eprint{hep-ph/9701219}.

\bibinfo{author}{\bibfnamefont{A.~B.} \bibnamefont{Lahanas}},
  \bibinfo{author}{\bibfnamefont{D.~V.} \bibnamefont{Nanopoulos}},
  \bibnamefont{and} \bibinfo{author}{\bibfnamefont{V.~C.}
  \bibnamefont{Spanos}}, \bibinfo{journal}{Mod. Phys. Lett.}
  \textbf{\bibinfo{volume}{A16}}, \bibinfo{pages}{1229} (\bibinfo{year}{2001}),
  \eprint{hep-ph/0009065}.

\bibinfo{author}{\bibfnamefont{D.~G.} \bibnamefont{Cerdeno}} \bibnamefont{and}
  \bibinfo{author}{\bibfnamefont{C.}~\bibnamefont{Munoz}},
  \bibinfo{journal}{JHEP} \textbf{\bibinfo{volume}{10}}, \bibinfo{pages}{015}
  (\bibinfo{year}{2004}), \eprint{hep-ph/0405057}.

\bibinfo{author}{\bibfnamefont{A.}~\bibnamefont{Masiero}},
  \bibinfo{author}{\bibfnamefont{S.}~\bibnamefont{Profumo}}, \bibnamefont{and}
  \bibinfo{author}{\bibfnamefont{P.}~\bibnamefont{Ullio}},
  \bibinfo{journal}{Nucl. Phys.} \textbf{\bibinfo{volume}{B712}},
  \bibinfo{pages}{86} (\bibinfo{year}{2005}), \eprint{hep-ph/0412058}.

\bibitem[{\citenamefont{Griest and Seckel}(1991)}]{Griest:1990kh}
\bibinfo{author}{\bibfnamefont{K.}~\bibnamefont{Griest}} \bibnamefont{and}
  \bibinfo{author}{\bibfnamefont{D.}~\bibnamefont{Seckel}},
  \bibinfo{journal}{Phys. Rev.} \textbf{\bibinfo{volume}{D43}},
  \bibinfo{pages}{3191} (\bibinfo{year}{1991}).

\bibitem[{\citenamefont{Lee and Weinberg}(1977)}]{Lee:1977ua}
\bibinfo{author}{\bibfnamefont{B.~W.} \bibnamefont{Lee}} \bibnamefont{and}
  \bibinfo{author}{\bibfnamefont{S.}~\bibnamefont{Weinberg}},
  \bibinfo{journal}{Phys. Rev. Lett.} \textbf{\bibinfo{volume}{39}},
  \bibinfo{pages}{165} (\bibinfo{year}{1977}).

\bibitem[{\citenamefont{Hisano et~al.}(2004)\citenamefont{Hisano, Matsumoto,
  and Nojiri}}]{Hisano:2003ec}
\bibinfo{author}{\bibfnamefont{J.}~\bibnamefont{Hisano}},
  \bibinfo{author}{\bibfnamefont{S.}~\bibnamefont{Matsumoto}},
  \bibnamefont{and} \bibinfo{author}{\bibfnamefont{M.~M.}
  \bibnamefont{Nojiri}}, \bibinfo{journal}{Phys. Rev. Lett.}
  \textbf{\bibinfo{volume}{92}}, \bibinfo{pages}{031303}
  (\bibinfo{year}{2004}), \eprint{hep-ph/0307216}.

\bibitem[{\citenamefont{Matsumoto et~al.}(2005)\citenamefont{Matsumoto, Sato,
  and Sato}}]{Matsumoto:2005ui}
\bibinfo{author}{\bibfnamefont{S.}~\bibnamefont{Matsumoto}},
  \bibinfo{author}{\bibfnamefont{J.}~\bibnamefont{Sato}}, \bibnamefont{and}
  \bibinfo{author}{\bibfnamefont{Y.}~\bibnamefont{Sato}}
  (\bibinfo{year}{2005}), \eprint{hep-ph/0505160}.

\bibitem[{\citenamefont{Profumo et~al.}(2005)\citenamefont{Profumo, Sigurdson,
  Ullio, and Kamionkowski}}]{Profumo:2004qt}
\bibinfo{author}{\bibfnamefont{S.}~\bibnamefont{Profumo}},
  \bibinfo{author}{\bibfnamefont{K.}~\bibnamefont{Sigurdson}},
  \bibinfo{author}{\bibfnamefont{P.}~\bibnamefont{Ullio}}, \bibnamefont{and}
  \bibinfo{author}{\bibfnamefont{M.}~\bibnamefont{Kamionkowski}},
  \bibinfo{journal}{Phys. Rev.} \textbf{\bibinfo{volume}{D71}},
  \bibinfo{pages}{023518} (\bibinfo{year}{2005}), \eprint{astro-ph/0410714}.

\bibitem[{\citenamefont{Gladyshev et~al.}(2005)\citenamefont{Gladyshev,
  Kazakov, and Paucar}}]{Gladyshev:2005mn}
\bibinfo{author}{\bibfnamefont{A.~V.} \bibnamefont{Gladyshev}},
  \bibinfo{author}{\bibfnamefont{D.~I.} \bibnamefont{Kazakov}},
  \bibnamefont{and} \bibinfo{author}{\bibfnamefont{M.~G.} \bibnamefont{Paucar}}
  (\bibinfo{year}{2005}), \eprint{hep-ph/0509168}.

\bibitem[{\citenamefont{Feng et~al.}(2004{\natexlab{a}})\citenamefont{Feng, Su,
  and Takayama}}]{Feng:2004mt}
\bibinfo{author}{\bibfnamefont{J.~L.} \bibnamefont{Feng}},
  \bibinfo{author}{\bibfnamefont{S.}~\bibnamefont{Su}}, \bibnamefont{and}
  \bibinfo{author}{\bibfnamefont{F.}~\bibnamefont{Takayama}},
  \bibinfo{journal}{Phys. Rev.} \textbf{\bibinfo{volume}{D70}},
  \bibinfo{pages}{075019} (\bibinfo{year}{2004}{\natexlab{a}}),
  \eprint{hep-ph/0404231}.

\bibinfo{author}{\bibfnamefont{J.~L.} \bibnamefont{Feng}},
  \bibinfo{author}{\bibfnamefont{S.-f.} \bibnamefont{Su}}, \bibnamefont{and}
  \bibinfo{author}{\bibfnamefont{F.}~\bibnamefont{Takayama}},
  \bibinfo{journal}{Phys. Rev.} \textbf{\bibinfo{volume}{D70}},
  \bibinfo{pages}{063514} (\bibinfo{year}{2004}{\natexlab{b}}),
  \eprint{hep-ph/0404198}.

\bibinfo{author}{\bibfnamefont{J.~L.} \bibnamefont{Feng}},
  \bibinfo{author}{\bibfnamefont{A.}~\bibnamefont{Rajaraman}},
  \bibnamefont{and} \bibinfo{author}{\bibfnamefont{F.}~\bibnamefont{Takayama}},
  \bibinfo{journal}{Phys. Rev. Lett.} \textbf{\bibinfo{volume}{91}},
  \bibinfo{pages}{011302} (\bibinfo{year}{2003}), \eprint{hep-ph/0302215}.

\bibinfo{author}{\bibfnamefont{J.~L.} \bibnamefont{Feng}} \bibnamefont{and}
  \bibinfo{author}{\bibfnamefont{T.}~\bibnamefont{Moroi}},
  \bibinfo{journal}{Phys. Rev.} \textbf{\bibinfo{volume}{D58}},
  \bibinfo{pages}{035001} (\bibinfo{year}{1998}), \eprint{hep-ph/9712499}.

\bibinfo{author}{\bibfnamefont{W.}~\bibnamefont{Buchmuller}},
  \bibinfo{author}{\bibfnamefont{K.}~\bibnamefont{Hamaguchi}},
  \bibinfo{author}{\bibfnamefont{M.}~\bibnamefont{Ratz}}, \bibnamefont{and}
  \bibinfo{author}{\bibfnamefont{T.}~\bibnamefont{Yanagida}},
  \bibinfo{journal}{Phys. Lett.} \textbf{\bibinfo{volume}{B588}},
  \bibinfo{pages}{90} (\bibinfo{year}{2004}), \eprint{hep-ph/0402179}.

\bibinfo{author}{\bibfnamefont{W.}~\bibnamefont{Buchmuller}},
  \bibinfo{author}{\bibfnamefont{K.}~\bibnamefont{Hamaguchi}},
  \bibnamefont{and} \bibinfo{author}{\bibfnamefont{M.}~\bibnamefont{Ratz}},
  \bibinfo{journal}{Phys. Lett.} \textbf{\bibinfo{volume}{B574}},
  \bibinfo{pages}{156} (\bibinfo{year}{2003}), \eprint{hep-ph/0307181}.

\bibinfo{author}{\bibfnamefont{W.}~\bibnamefont{Buchmuller}},
  \bibinfo{author}{\bibfnamefont{K.}~\bibnamefont{Hamaguchi}},
  \bibnamefont{and} \bibinfo{author}{\bibfnamefont{J.}~\bibnamefont{Kersten}},
  \bibinfo{journal}{Phys. Lett.} \textbf{\bibinfo{volume}{B632}},
  \bibinfo{pages}{366} (\bibinfo{year}{2006}), \eprint{hep-ph/0506105}.

\bibinfo{author}{\bibfnamefont{W.}~\bibnamefont{Buchmuller}},
  \bibinfo{author}{\bibfnamefont{J.}~\bibnamefont{Kersten}}, \bibnamefont{and}
  \bibinfo{author}{\bibfnamefont{K.}~\bibnamefont{Schmidt-Hoberg}}
  (\bibinfo{year}{2005}), \eprint{hep-ph/0512152}.

\bibinfo{author}{\bibfnamefont{F.}~\bibnamefont{Takayama}} \bibnamefont{and}
  \bibinfo{author}{\bibfnamefont{M.}~\bibnamefont{Yamaguchi}},
  \bibinfo{journal}{Phys. Lett.} \textbf{\bibinfo{volume}{B485}},
  \bibinfo{pages}{388} (\bibinfo{year}{2000}), \eprint{hep-ph/0005214}.

\bibinfo{author}{\bibfnamefont{D.~J.} \bibnamefont{Muller}} \bibnamefont{and}
  \bibinfo{author}{\bibfnamefont{S.}~\bibnamefont{Nandi}},
  \bibinfo{journal}{Phys. Rev.} \textbf{\bibinfo{volume}{D60}},
  \bibinfo{pages}{015008} (\bibinfo{year}{1999}), \eprint{hep-ph/9811248}.

\bibinfo{author}{\bibfnamefont{S.}~\bibnamefont{Ambrosanio}}
  \bibnamefont{et~al.} (\bibinfo{year}{2000}), \eprint{hep-ph/0012192}.

\bibinfo{author}{\bibfnamefont{M.}~\bibnamefont{Hirsch}},
  \bibinfo{author}{\bibfnamefont{W.}~\bibnamefont{Porod}}, \bibnamefont{and}
  \bibinfo{author}{\bibfnamefont{D.}~\bibnamefont{Restrepo}},
  \bibinfo{journal}{JHEP} \textbf{\bibinfo{volume}{03}}, \bibinfo{pages}{062}
  (\bibinfo{year}{2005}), \eprint{hep-ph/0503059}.

\bibinfo{author}{\bibfnamefont{K.}~\bibnamefont{Jedamzik}},
  \bibinfo{author}{\bibfnamefont{M.}~\bibnamefont{Lemoine}}, \bibnamefont{and}
  \bibinfo{author}{\bibfnamefont{G.}~\bibnamefont{Moultaka}}
  (\bibinfo{year}{2005}{\natexlab{a}}), \eprint{astro-ph/0508141}.

\bibinfo{author}{\bibfnamefont{F.}~\bibnamefont{Wang}} \bibnamefont{and}
  \bibinfo{author}{\bibfnamefont{J.~M.} \bibnamefont{Yang}},
  \bibinfo{journal}{Nucl. Phys.} \textbf{\bibinfo{volume}{B709}},
  \bibinfo{pages}{409} (\bibinfo{year}{2005}), \eprint{hep-ph/0408271}.

\bibinfo{author}{\bibfnamefont{D.~G.} \bibnamefont{Cerdeno}},
  \bibinfo{author}{\bibfnamefont{K.-Y.} \bibnamefont{Choi}},
  \bibinfo{author}{\bibfnamefont{K.}~\bibnamefont{Jedamzik}},
  \bibinfo{author}{\bibfnamefont{L.}~\bibnamefont{Roszkowski}},
  \bibnamefont{and} \bibinfo{author}{\bibfnamefont{R.}~\bibnamefont{Ruiz~de
  Austri}} (\bibinfo{year}{2005}), \eprint{hep-ph/0509275}.

\bibinfo{author}{\bibfnamefont{L.}~\bibnamefont{Roszkowski}},
  \bibinfo{author}{\bibfnamefont{R.}~\bibnamefont{Ruiz~de Austri}},
  \bibnamefont{and} \bibinfo{author}{\bibfnamefont{K.-Y.} \bibnamefont{Choi}},
  \bibinfo{journal}{JHEP} \textbf{\bibinfo{volume}{08}}, \bibinfo{pages}{080}
  (\bibinfo{year}{2005}), \eprint{hep-ph/0408227}.

\bibinfo{author}{\bibfnamefont{F.}~\bibnamefont{Wang}} \bibnamefont{and}
  \bibinfo{author}{\bibfnamefont{J.~M.} \bibnamefont{Yang}},
  \bibinfo{journal}{Eur. Phys. J.} \textbf{\bibinfo{volume}{C38}},
  \bibinfo{pages}{129} (\bibinfo{year}{2004}), \eprint{hep-ph/0405186}.

\bibinfo{author}{\bibfnamefont{P.~G.} \bibnamefont{Mercadante}},
  \bibinfo{author}{\bibfnamefont{J.~K.} \bibnamefont{Mizukoshi}},
  \bibnamefont{and} \bibinfo{author}{\bibfnamefont{H.}~\bibnamefont{Yamamoto}},
  \bibinfo{journal}{Phys. Rev.} \textbf{\bibinfo{volume}{D64}},
  \bibinfo{pages}{015005} (\bibinfo{year}{2001}), \eprint{hep-ph/0010067}.

\bibinfo{author}{\bibfnamefont{J.~L.} \bibnamefont{Feng}} \bibnamefont{and}
  \bibinfo{author}{\bibfnamefont{B.~T.} \bibnamefont{Smith}},
  \bibinfo{journal}{Phys. Rev.} \textbf{\bibinfo{volume}{D71}},
  \bibinfo{pages}{015004} (\bibinfo{year}{2005}), \eprint{hep-ph/0409278}.

\bibinfo{author}{\bibfnamefont{M.}~\bibnamefont{Kaplinghat}},
  \bibinfo{journal}{Phys. Rev.} \textbf{\bibinfo{volume}{D72}},
  \bibinfo{pages}{063510} (\bibinfo{year}{2005}), \eprint{astro-ph/0507300}.

\bibinfo{author}{\bibfnamefont{K.-Y.} \bibnamefont{Choi}} \bibnamefont{and}
  \bibinfo{author}{\bibfnamefont{L.}~\bibnamefont{Roszkowski}},
  \bibinfo{journal}{AIP Conf. Proc.} \textbf{\bibinfo{volume}{805}},
  \bibinfo{pages}{30} (\bibinfo{year}{2006}), \eprint{hep-ph/0511003}.

\bibinfo{author}{\bibfnamefont{O.}~\bibnamefont{Seto}} (\bibinfo{year}{2005}),
  \eprint{hep-ph/0512071}.

\bibinfo{author}{\bibfnamefont{F.}~\bibnamefont{Wang}},
  \bibinfo{author}{\bibfnamefont{W.}~\bibnamefont{Wang}}, \bibnamefont{and}
  \bibinfo{author}{\bibfnamefont{J.~M.} \bibnamefont{Yang}},
  \bibinfo{journal}{Phys. Rev.} \textbf{\bibinfo{volume}{D72}},
  \bibinfo{pages}{077701} (\bibinfo{year}{2005}), \eprint{hep-ph/0507172}.

\bibinfo{author}{\bibfnamefont{A.}~\bibnamefont{Brandenburg}},
  \bibinfo{author}{\bibfnamefont{L.}~\bibnamefont{Covi}},
  \bibinfo{author}{\bibfnamefont{K.}~\bibnamefont{Hamaguchi}},
  \bibinfo{author}{\bibfnamefont{L.}~\bibnamefont{Roszkowski}},
  \bibnamefont{and} \bibinfo{author}{\bibfnamefont{F.~D.}
  \bibnamefont{Steffen}}, \bibinfo{journal}{Phys. Lett.}
  \textbf{\bibinfo{volume}{B617}}, \bibinfo{pages}{99} (\bibinfo{year}{2005}),
  \eprint{hep-ph/0501287}.

\bibinfo{author}{\bibfnamefont{M.}~\bibnamefont{Ibe}} \bibnamefont{and}
  \bibinfo{author}{\bibfnamefont{T.}~\bibnamefont{Yanagida}},
  \bibinfo{journal}{Phys. Lett.} \textbf{\bibinfo{volume}{B597}},
  \bibinfo{pages}{47} (\bibinfo{year}{2004}), \eprint{hep-ph/0404134}.

\bibinfo{author}{\bibfnamefont{K.}~\bibnamefont{Jedamzik}},
  \bibinfo{author}{\bibfnamefont{M.}~\bibnamefont{Lemoine}}, \bibnamefont{and}
  \bibinfo{author}{\bibfnamefont{G.}~\bibnamefont{Moultaka}}
  (\bibinfo{year}{2005}{\natexlab{b}}), \eprint{hep-ph/0506129}.

\bibinfo{author}{\bibfnamefont{A.}~\bibnamefont{De~Roeck}} \bibnamefont{et~al.}
  (\bibinfo{year}{2005}), \eprint{hep-ph/0508198}.

\bibinfo{author}{\bibfnamefont{M.}~\bibnamefont{Bolz}},
  \bibinfo{author}{\bibfnamefont{A.}~\bibnamefont{Brandenburg}},
  \bibnamefont{and}
  \bibinfo{author}{\bibfnamefont{W.}~\bibnamefont{Buchmuller}},
  \bibinfo{journal}{Nucl. Phys.} \textbf{\bibinfo{volume}{B606}},
  \bibinfo{pages}{518} (\bibinfo{year}{2001}), \eprint{hep-ph/0012052}.






\bibitem[{\citenamefont{Hamaguchi and Ibarra}(2005)}]{Hamaguchi:2004ne}
\bibinfo{author}{\bibfnamefont{K.}~\bibnamefont{Hamaguchi}} \bibnamefont{and}
  \bibinfo{author}{\bibfnamefont{A.}~\bibnamefont{Ibarra}},
  \bibinfo{journal}{JHEP} \textbf{\bibinfo{volume}{02}}, \bibinfo{pages}{028}
  (\bibinfo{year}{2005}), \eprint{hep-ph/0412229}.

\bibitem[{\citenamefont{Bi et~al.}(2004)\citenamefont{Bi, Wang, Zhang, and
  Zhang}}]{Bi:2004ys}
\bibinfo{author}{\bibfnamefont{X.-J.} \bibnamefont{Bi}},
  \bibinfo{author}{\bibfnamefont{J.-X.} \bibnamefont{Wang}},
  \bibinfo{author}{\bibfnamefont{C.}~\bibnamefont{Zhang}}, \bibnamefont{and}
  \bibinfo{author}{\bibfnamefont{X.-m.} \bibnamefont{Zhang}},
  \bibinfo{journal}{Phys. Rev.} \textbf{\bibinfo{volume}{D70}},
  \bibinfo{pages}{123512} (\bibinfo{year}{2004}), \eprint{hep-ph/0404263}.

\bibinfo{author}{\bibfnamefont{D.}~\bibnamefont{Hooper}} \bibnamefont{and}
  \bibinfo{author}{\bibfnamefont{L.-T.} \bibnamefont{Wang}},
  \bibinfo{journal}{Phys. Rev.} \textbf{\bibinfo{volume}{D70}},
  \bibinfo{pages}{063506} (\bibinfo{year}{2004}), \eprint{hep-ph/0402220}.

\bibinfo{author}{\bibfnamefont{H.-B.} \bibnamefont{Kim}} \bibnamefont{and}
  \bibinfo{author}{\bibfnamefont{J.~E.} \bibnamefont{Kim}},
  \bibinfo{journal}{Phys. Lett.} \textbf{\bibinfo{volume}{B527}},
  \bibinfo{pages}{18} (\bibinfo{year}{2002}), \eprint{hep-ph/0108101}.

\bibinfo{author}{\bibfnamefont{A.}~\bibnamefont{Brandenburg}} \bibnamefont{and}
  \bibinfo{author}{\bibfnamefont{F.~D.} \bibnamefont{Steffen}},
  \bibinfo{journal}{JCAP} \textbf{\bibinfo{volume}{0408}}, \bibinfo{pages}{008}
  (\bibinfo{year}{2004}), \eprint{hep-ph/0405158}.

\bibitem[{\citenamefont{Hamaguchi et~al.}(2004)\citenamefont{Hamaguchi, Kuno,
  Nakaya, and Nojiri}}]{Hamaguchi:2004df}
\bibinfo{author}{\bibfnamefont{K.}~\bibnamefont{Hamaguchi}},
  \bibinfo{author}{\bibfnamefont{Y.}~\bibnamefont{Kuno}},
  \bibinfo{author}{\bibfnamefont{T.}~\bibnamefont{Nakaya}}, \bibnamefont{and}
  \bibinfo{author}{\bibfnamefont{M.~M.} \bibnamefont{Nojiri}},
  \bibinfo{journal}{Phys. Rev.} \textbf{\bibinfo{volume}{D70}},
  \bibinfo{pages}{115007} (\bibinfo{year}{2004}), \eprint{hep-ph/0409248}.

\bibitem[{\citenamefont{Feng and Smith}(2005)}]{Feng:2004yi}
\bibinfo{author}{\bibfnamefont{J.~L.} \bibnamefont{Feng}} \bibnamefont{and}
  \bibinfo{author}{\bibfnamefont{B.~T.} \bibnamefont{Smith}},
  \bibinfo{journal}{Phys. Rev.} \textbf{\bibinfo{volume}{D71}},
  \bibinfo{pages}{015004} (\bibinfo{year}{2005}), \eprint{hep-ph/0409278}.

\bibitem[{\citenamefont{Jittoh et~al.}(2005)\citenamefont{Jittoh, Matsumoto,
  Sato, Sato, and Takeda}}]{Jittoh:2004bz}
\bibinfo{author}{\bibfnamefont{T.}~\bibnamefont{Jittoh}},
  \bibinfo{author}{\bibfnamefont{S.}~\bibnamefont{Matsumoto}},
  \bibinfo{author}{\bibfnamefont{J.}~\bibnamefont{Sato}},
  \bibinfo{author}{\bibfnamefont{Y.}~\bibnamefont{Sato}}, \bibnamefont{and}
  \bibinfo{author}{\bibfnamefont{K.}~\bibnamefont{Takeda}},
  \bibinfo{journal}{Phys. Rev.} \textbf{\bibinfo{volume}{A71}},
  \bibinfo{pages}{012109} (\bibinfo{year}{2005}), \eprint{quant-ph/0408149}.

\bibitem[{\citenamefont{Kawasaki et~al.}(2001)\citenamefont{Kawasaki, Kohri,
  and Moroi}}]{Kawasaki:2000qr}
\bibinfo{author}{\bibfnamefont{M.}~\bibnamefont{Kawasaki}},
  \bibinfo{author}{\bibfnamefont{K.}~\bibnamefont{Kohri}}, \bibnamefont{and}
  \bibinfo{author}{\bibfnamefont{T.}~\bibnamefont{Moroi}},
  \bibinfo{journal}{Phys. Rev.} \textbf{\bibinfo{volume}{D63}},
  \bibinfo{pages}{103502} (\bibinfo{year}{2001}), \eprint{hep-ph/0012279}.

\bibitem[{\citenamefont{Kawasaki et~al.}(2005)\citenamefont{Kawasaki, Kohri,
  and Moroi}}]{Kawasaki:2004qu}
\bibinfo{author}{\bibfnamefont{M.}~\bibnamefont{Kawasaki}},
  \bibinfo{author}{\bibfnamefont{K.}~\bibnamefont{Kohri}}, \bibnamefont{and}
  \bibinfo{author}{\bibfnamefont{T.}~\bibnamefont{Moroi}},
  \bibinfo{journal}{Phys. Rev.} \textbf{\bibinfo{volume}{D71}},
  \bibinfo{pages}{083502} (\bibinfo{year}{2005}), \eprint{astro-ph/0408426}.

\bibitem[{\citenamefont{Hisano and Nomura}(1999)}]{Hisano:1998fj}
\bibinfo{author}{\bibfnamefont{J.}~\bibnamefont{Hisano}} \bibnamefont{and}
  \bibinfo{author}{\bibfnamefont{D.}~\bibnamefont{Nomura}},
  \bibinfo{journal}{Phys. Rev.} \textbf{\bibinfo{volume}{D59}},
  \bibinfo{pages}{116005} (\bibinfo{year}{1999}), \eprint{hep-ph/9810479}.

\bibitem[{\citenamefont{Hisano et~al.}(1996)\citenamefont{Hisano, Moroi, Tobe,
  and Yamaguchi}}]{Hisano:1995cp}
\bibinfo{author}{\bibfnamefont{J.}~\bibnamefont{Hisano}},
  \bibinfo{author}{\bibfnamefont{T.}~\bibnamefont{Moroi}},
  \bibinfo{author}{\bibfnamefont{K.}~\bibnamefont{Tobe}}, \bibnamefont{and}
  \bibinfo{author}{\bibfnamefont{M.}~\bibnamefont{Yamaguchi}},
  \bibinfo{journal}{Phys. Rev.} \textbf{\bibinfo{volume}{D53}},
  \bibinfo{pages}{2442} (\bibinfo{year}{1996}), \eprint{hep-ph/9510309}.

\bibitem[{\citenamefont{Hisano et~al.}(1995)\citenamefont{Hisano, Moroi, Tobe,
  Yamaguchi, and Yanagida}}]{Hisano:1995nq}
\bibinfo{author}{\bibfnamefont{J.}~\bibnamefont{Hisano}},
  \bibinfo{author}{\bibfnamefont{T.}~\bibnamefont{Moroi}},
  \bibinfo{author}{\bibfnamefont{K.}~\bibnamefont{Tobe}},
  \bibinfo{author}{\bibfnamefont{M.}~\bibnamefont{Yamaguchi}},
  \bibnamefont{and} \bibinfo{author}{\bibfnamefont{T.}~\bibnamefont{Yanagida}},
  \bibinfo{journal}{Phys. Lett.} \textbf{\bibinfo{volume}{B357}},
  \bibinfo{pages}{579} (\bibinfo{year}{1995}), \eprint{hep-ph/9501407}.

\bibitem[{\citenamefont{Sato and Tobe}(2001)}]{Sato:2000ff}
\bibinfo{author}{\bibfnamefont{J.}~\bibnamefont{Sato}} \bibnamefont{and}
  \bibinfo{author}{\bibfnamefont{K.}~\bibnamefont{Tobe}},
  \bibinfo{journal}{Phys. Rev.} \textbf{\bibinfo{volume}{D63}},
  \bibinfo{pages}{116010} (\bibinfo{year}{2001}), \eprint{hep-ph/0012333}.

\bibitem[{\citenamefont{Ota and Sato}(2005)}]{Ota:2005et}
\bibinfo{author}{\bibfnamefont{T.}~\bibnamefont{Ota}} \bibnamefont{and}
  \bibinfo{author}{\bibfnamefont{J.}~\bibnamefont{Sato}},
  \bibinfo{journal}{Phys. Rev.} \textbf{\bibinfo{volume}{D71}},
  \bibinfo{pages}{096004} (\bibinfo{year}{2005}), \eprint{hep-ph/0502124}.

\end{thebibliography}

\end{document}